# Auto-Regressive Control of Execution Costs

Vidyadhar Kulkarni & Simeon Kolev

RTG: Foundations in Probability, Optimization and Data Sciences

UNC Spring 2024

**Abstract**

Bertsimas and Lo's seminal work established a foundational framework for addressing the implementation shortfall dilemma faced by large institutional investors. Their models emphasized the critical role of accurate knowledge of market microstructure and price/information dynamics in optimizing trades to minimize execution costs. However, this paper recognizes that perfect initial knowledge may not be a realistic assumption for new investors entering the market. Therefore, this study aims to bridge this gap by proposing an approach that iteratively derives OLS estimates of the market parameters from period to period. This methodology enables uninformed investors to engage in the market dynamically, adjusting their strategies over time based on evolving estimates, thus offering a practical solution for navigating the complexities of execution cost optimization without perfect initial knowledge.

# Contents

**Introduction**

Large Institutional Investor face particular problems when it comes to the measurement and management of profits and trading costs. The incomprehensible amount of assets that mutual, pension, and hedge funds manage gives them the privilege of reliable sustenance at a risk-free rate of returns. For this reason, it is often, that large asset managers prioritize efficient trade execution to keep as much of their earnings as possible. Many large firms have found that a shift in emphasis in the management execution costs (commission, bid/ask spread, and order size costs) have had a more substantial impact on retained earnings than strategies prioritizing marginal improvements in trade yields. Analogously, many asset managers have found that "implementation shortfall" is a surprisingly large hinderance to portfolio growth, underscoring the importance of the optimal control of execution costs.

The Optimal Control of Execution Costs, by Bertsimas and Lo at MIT laid the foundation for many price impact models that aim to mitigate excess execution costs. The objective when mitigating implementation shortfall is to make the best trades given the nature of current and future market dynamics as described by the models. Defining and controlling executions costs are fundamentally dynamic optimization problems and must be solved as such. To optimize execution costs, the dynamic strategy must take into consideration current and future prices and execute trades in such a manner as to minimize total expected cost. The strategies derived from these models aim to exploit market dynamics and minimize the expected costs of executing large trades over a fixed horizon. Specifically, given a fixed block of shares ($S_t$) to be executed within a fixed finite number of periods ($T$).

The general price impact models analyzed by Bertsimas and Lo are variations of additive permanent price impact models. The price models serve as discrete approximations of a stock's price ($P_t$) given, order size ($B_t$), exogenous variables ($X_t$), time-series correlations ($\rho$), parameter relevance ($\theta, \gamma$), and random fluctuations ($\varepsilon_t, \eta_t$). The optimal ($B_t$) strategies are derived by minimizing the Bellman equation ($\sum_1^T P_t B_t$) and solving recursively through stochastic dynamic programing to obtain an explicit closed form expression of the "best-execution strategy."

In this paper, I will be exploring the additive price-impact models proposed by Bertsimas and Lo. After reviewing the literature, using their established and tested models, I will be building upon the optimal "Informed" price-impact model by providing an alternative Auto-Regressive price-impact model. I intend to analyze the importance of the true model parameters when implementing the optimal execution strategy in both models. Additionally, I will be analyzing how the optimal strategy and minimal execution costs deviate from their theoretical expectations in the hypothetical that a uniformed investor enters the market and is forced to derive their own best-estimates of the true market dynamics. Specially, aiming to estimate the excess costs associated with making suboptimal buying decisions until a reasonable estimate of the market parameters is derived.

The results in this paper will provide simulated estimates of the per share improvement and variance in cumulative execution costs found under the framework of the following 3 models, specifically focusing on the performance of the "Auto-Regressive" strategy. To quantify the importance of the true market parameters, the performance of the following models must be evaluated in a real-world context with consideration of the decisions institutional investors will likely make when analyzing the cost-benefit of entering the market informed vs uninformed.

The summaries of the papers below reemphasize the importance of market information, algorithmic strategies, and ability to control stochastic market dynamics for favorable execution outcomes.

## Literature Review

"Optimal Control of Execution Costs" by Dimitris Bertsimas and Andrew W. Lo (1998). Bertsimas and Lo's seminal work revolutionized execution cost optimization by introducing additive permanent price impact models. Their research identified that optimal execution strategies must account for both current and future price dynamics to minimize execution costs effectively. They introduced a framework grounded in stochastic dynamic programming, which offers a comprehensive approach to optimizing execution strategies over time. By analyzing various factors such as order size, exogenous variables, and time-series correlations, they derived closed-form expressions of the "best-execution strategy." This framework serves as a cornerstone for subsequent research in execution cost optimization, facilitating the development of simulation models that accurately capture the complexities of trading dynamics.

"High-Frequency Trading in a Limit Order Book' by Marco Avellaneda and Sasha Stoikov (2008). Avellaneda and Stoikov's study focus on the nonlinear relationship between order size and execution costs, emphasizing the importance of optimizing execution strategies with respect to non-linear price-impacts to minimize market impact. Their research demonstrates that breaking down large trades into smaller ones can mitigate adverse price movements and achieve more favorable execution outcomes in the long run. By analyzing the trading profile of large orders, investors gain valuable insights into order execution dynamics. These insights inform the development of simulation models that accurately simulate order execution using real-world limit-order book distributions, facilitating the creation of pragmatic optimal execution strategies that minimize market impact and reduce execution costs.

"Optimal Trading with Stochastic Liquidity and Volatility" by Robert Almgren (2012). Almgrens contribution extends traditional execution cost optimization frameworks by incorporating stochastic liquidity and volatility dynamics. His research highlights the significance of adaptability in execution strategies to navigate uncertain market conditions effectively. By integrating stochastic elements into execution cost optimization frameworks, investors can assess the impact of varying market conditions on execution costs and implement optimal strategies conditional on varying degrees of risk aversion. The numerical analysis highlights the need for flexible execution strategies that can adjust to dynamic market environments. This adaptability enhances the resilience and effectiveness of execution cost optimization techniques, ensuring optimal performance under changing market and liquidity conditions.

"Machine Learning for Market Microstructure and High Frequency Trading" by Michael Kearns, Yuriy Nevmyvaka (2013). Kearns and Nevmyvaka research showcase the effectiveness of algorithmic trading strategies in execution cost optimization. Their findings demonstrate how machine learning can adaptively adjust execution strategies by predicting directional movements in volatile market conditions, and thereby minimizing costs and enhancing execution performance. By integrating algorithmic trading algorithms into simulation models, investors gain insights into the comparative effectiveness of automated trading approaches against traditional execution methods. This comparison enables the development of more efficient and cost-effective trading strategies, with a powerful and principled framework for trading optimization via historical data.

"High Frequency Trading and Price Discovery" by Jonathan Brogaard, Terrence Hendershott, and Ryan Riordan (2014). Brogaard, Hendershott, and Riordan analysis of the role of high-frequency traders (HTFs) in price discovery and price efficiency. In their study they reemphasis the importance of incorporating information in asset pricing and explain how the increase in HTFs merits the necessity for consideration of high-frequency transaction level data. Their finding suggests an increase in model complexity to produce high-frequency optimal strategies with the incorporation of market information, macro news announcements, market wide price movements, and limit order book imbalances. Their simulations show tendencies for HFTs to trade in the direction of permanent price chances and their information advances is sufficient to overcome big-ask spread, trading fees, to achieve favorable executions costs.

The following models and results, initially derived by Dimitris Bertsimas and Andrew W. Lo (1998) in the "Optimal Control of Execution Costs", including the strategies found solving the Bellman equation recursively using stochastic dynamic programming are summarized below. These are the models I intend to analyze by simulating optimal market strategies and subsequent execution costs under the permanent additive price-impact framework with the incorporation of serially correlated market information.

# The Naive Model

Suppose UNC is looking to buy 100,000 shares ($S_0$) of stock in ABC Inc, hypothetically trading at \$50 ($P_0$) a share. How should UNC approach buying ABC's stock so that they acquire ABC shares at the lowest aggregate cost possible? A financially irresponsible strategy would be to purchase all $S_0$ shares in a single market order. As short-term demand curves are relatively inelastic, UNC's market worder would be executed at the highest bid price available for each share. In this case, the optimal strategy would be to buy ABC stock into the smallest possible uniform batches at each period. The later strategy is found to be the Naive strategy, a best-execution strategy found assuming knowledge of the market order book distribution.

The Naive Model, more technically known as Additive Permanent Price Impact, is such that the price impact of market orders is linear and permanent. In this model there are 2 components, the price-impact from other investors' market orders ($\varepsilon_t$), and the price-impact that our market orders have on execution price ($\theta B_t$). The ($\varepsilon_t$) component is given by an arithmetic random walk, and the execution-impact is a linear function of order size ($B_t$).

The following equation summarizes the Additive Permanent Price Impact Model:

$$P_t = P_{t-1} + \theta B_t + \varepsilon_t$$

$$\varepsilon_t \sim N(0, \sigma^2)$$

The objective is to find the $B_t$ producing the minimal execution cost strategy given by the Bellman equation:

$$V_t = \min_{B_t} \sum_t^T P_t B_t$$

Solving the Bellman equation recursively yields the optimal order size and execution as:

$$B_t = \frac{S_t}{T - t + 1}$$

$$V_t = S_t P_t + \theta \frac{S_t^2}{2}\left(\frac{T - t + 2}{T - t + 1}\right)$$

Further inspection of the optimal of order size shows its simplification to a constant, $B_t = \frac{S}{T}$, meaning the optimal strategy under the Naive model is to uniformly split shares ($S_0$) over ($T$) periods.

Naive Numerical Example:

Suppose UNC is subject to the restraint of having to sell 100,000 shares ($S_0$) in 20 market orders ($T$), then, given $\theta = 5 \times 10^{-5}$ and $\sigma_\varepsilon^2 = (0.125)^2$, the Naive strategy suggests an optimal split of $B_t = 5,000$, and $V_t = \$5,262,500$.

Table 1 gives a simulated example of the performance of the Naive strategy.

| Period | Price | Shares.Bought | Shares.Remaining | Accumulated.Cost |
|---|---|---|---|---|
| 0 | 50.00000 | 0 | 100000 | 0.0 |
| 1 | 50.17994 | 5000 | 95000 | 250899.7 |
| 2 | 50.40117 | 5000 | 90000 | 502905.5 |
| 3 | 50.84601 | 5000 | 85000 | 757135.6 |
| 4 | 51.10482 | 5000 | 80000 | 1012659.7 |
| 5 | 51.37098 | 5000 | 75000 | 1269514.6 |
| 6 | 51.83536 | 5000 | 70000 | 1528691.4 |
| 7 | 52.14298 | 5000 | 65000 | 1789406.3 |
| 8 | 52.23485 | 5000 | 60000 | 2050580.5 |
| 9 | 52.39899 | 5000 | 55000 | 2312575.5 |
| 10 | 52.59328 | 5000 | 50000 | 2575541.9 |
| 11 | 52.99629 | 5000 | 45000 | 2840523.4 |
| 12 | 53.29127 | 5000 | 40000 | 3106979.7 |
| 13 | 53.59137 | 5000 | 35000 | 3374936.5 |
| 14 | 53.85520 | 5000 | 30000 | 3644212.5 |
| 15 | 54.03572 | 5000 | 25000 | 3914391.1 |
| 16 | 54.50908 | 5000 | 20000 | 4186936.6 |
| 17 | 54.82132 | 5000 | 15000 | 4461043.1 |
| 18 | 54.82549 | 5000 | 10000 | 4735170.6 |
| 19 | 55.16316 | 5000 | 5000 | 5010986.4 |
| 20 | 55.35406 | 5000 | 0 | 5287756.7 |

Actual Cost: 5287757
Expected Cost: 5262500
Improvement: -0.2525668

*Table 1*

The actual execution cost underperforms the theoretical optimal by -\$0.25 a share. Note, accumulated cost often differs from expected cost, as it is only when $\lim\limits_{T \to \infty} \frac{1}{T}\sum_0^T \varepsilon_t = 0$

The Naive model serves to illustrate that in the simplest scenario, in the absence of predictable exogenous variables correlated with price, the optimal strategy is irrespective of the price-impact from other investor's market orders ($\varepsilon_t$) and instead emphases uniformly splitting order size across ($T$) such as to minimize execution price-impact ($\Delta P_t$). However, the optimal strategy, under more complex price-impact functions which incorporate serially correlated exogenous variables, differs significantly from the Naive strategy. Thus, the best-execution strategy becomes a nontrivial function of the exogenous variables ($X_t$), resulting in dynamic splitting decisions.

## The Informed Model

Now suppose that UNC has access to some serially correlated data, such as interest rates, market conditions, economic trends, etc. How should UNC approach buying ABC's stock, while incorporating their data so that they acquire ABC shares at the lowest aggregate cost possible? In this scenario, the Naive strategy displays implementation shortfall, especially when market information is significantly predictable. As changes in market information influence price movements, UNC's uniform split will fail to capitalize in periods of discount and overcompensate in periods of premium, resulting in suboptimal execution costs. The new optimal strategy would be to buy ABC stock in correspondence to the changes in market information. With this approach, UNC will buy their largest quantity of shares in periods when $\left(\frac{dX}{dt} = 0\right)$ and $\left(\frac{d^2X}{dt^2} > 0\right)$, producing the lowest aggregate cost possible. This approach is found to be the Informed strategy, a best-execution strategy found assuming the availability of relevant serially correlated market information.

The Informed Model, more technically known as Additive Permanent Price Impact with Information, is such that the price-impact of market orders and changes in market information are linear and permanent. In this model there are 5 components, the price-impact from other investors' market orders $(\varepsilon_t)$, the price-impact that our market orders have on execution price $(\theta B_t)$, the price impact of serially correlated market information $(\gamma X_t)$, the information correlation coefficient $(\rho)$, and market information fluctuations $(\eta_t)$. The $(\varepsilon_t, \eta_t)$ components are given by an arithmetic random walk, and the execution-impact and information impact is an additive linear function of order size $(B_t)$ and information $(X_t)$.

The following equations summarizes the Additive Permanent Price Impact with Information:

$$P_t = P_{t-1} + \theta B_t + \gamma X_t + \varepsilon_t$$

$$X_t = \rho X_{t-1} + \eta_t$$

$$\varepsilon_t \sim N(0, \sigma_\varepsilon^2), \qquad \eta_t \sim N(0, \sigma_\eta^2)$$

The objective is to find the $B_t$ producing the minimal execution cost strategy given by the Bellman equation, and dynamic parameters:

$$V_t = \min_{B_t} \sum_t^T P_t B_t$$

$$a_t = \frac{\theta}{2}\left(1 + \frac{1}{T-t+1}\right), \qquad b_t = \gamma + \frac{\theta\rho}{2}\left(\frac{b_{t-1}}{a_{t-1}}\right),$$

$$c_t = \rho^2 c_{t-1} + \frac{\rho^2}{4}\left(\frac{b_{t-1}^2}{a_{t-1}}\right), \qquad d_t = d_{t-1} + c_{t-1}\sigma_\eta^2,$$

$$e_t = \frac{1}{T-t+1}, \qquad f_t = \frac{\rho}{2}\left(\frac{b_{t-1}}{a_{t-1}}\right)$$

Solving the Bellman equation by dynamic programming yields the optimal order size and execution as:

$$B_t = e_t S_t + f_t X_{t-1}$$

$$V_t = S_t P_t + a_t S_t^2 + b_t S_t X_t + c_t X_t^2 + d_t$$

Further inspection of the optimal order size shows its simplification to combination of the Naive constant and some adjustment factor, and $\mathrm{E}(B_t \,|\, X_{t-1}) = \frac{S_t}{T} + f_t X_{t-1}$, meaning the optimal strategy under the Informed model is to uniformly split shares $(S_0)$ over $(T)$ periods, and adjust by a factor of $f_t X_{t-1}$.

Informed Numerical Example:

Suppose UNC is subject to the restraint of having to sell 100,000 shares ($S_0$) in 20 market orders ($T$), then, given $\theta = 5 \times 10^{-5}$, $\gamma = 5$, $\rho = 0.5$, $\sigma_\epsilon^2 = (0.125)^2$, and $\sigma_\eta^2 = 0.001$ the Informed strategy suggests an optimal split of $B_t = 5{,}000 + f_t X_{t-1}$, and $V_t = \$5{,}258{,}727$.

Table 2 gives a simulated example of the performance of the Informed strategy.

| Period | Price | Shares.Bought | Shares.Remaining | Market.Information | Accumulated.Cost |
|---|---|---|---|---|---|
| 0 | 50.00000 | 0.0000 | 100000.000 | 0.00000000 | 0.0 |
| 1 | 50.41127 | 5000.0000 | 95000.000 | 0.06213838 | 252056.3 |
| 2 | 50.81902 | 6111.9501 | 88888.050 | 0.03668100 | 562659.7 |
| 3 | 51.17225 | 5593.8875 | 83294.162 | -0.14474626 | 8489115 |
| 4 | 50.85823 | 2345.3083 | 80948.854 | -0.27236529 | 968189.7 |
| 5 | 50.97548 | 292.9005 | 80655.954 | -0.17206901 | 9851205 |
| 6 | 51.18120 | 2394.5201 | 78261.433 | 0.03185352 | 11056749 |
| 7 | 51.36029 | 6136.1682 | 72125.265 | 0.06855009 | 14208303 |
| 8 | 51.43538 | 6708.2015 | 65417.064 | -0.14345136 | 1765869.2 |
| 9 | 51.41297 | 3060.4493 | 62356.614 | -0.18337476 | 1923216.0 |
| 10 | 51.23934 | 2667.7797 | 59688.835 | -0.20672216 | 2060178.0 |
| 11 | 50.79437 | 2660.5213 | 57028.313 | -0.39339165 | 2195617.5 |
| 12 | 50.63857 | 213.6387 | 56814.675 | -0.26115695 | 2206135.9 |
| 13 | 50.60266 | 3179.3794 | 53635.295 | -0.12633135 | 2367020.9 |
| 14 | 50.99780 | 5851.8117 | 47783.484 | 0.08097547 | 2665450.4 |
| 15 | 51.87965 | 9052.0218 | 38731.462 | 0.29890060 | 3134523.1 |
| 16 | 52.73807 | 11407.8247 | 27323.637 | 0.15904554 | 3736149.7 |
| 17 | 53.61813 | 8520.7681 | 18802.869 | 0.37164964 | 4193017.4 |
| 18 | 54.42512 | 9364.7033 | 9438.166 | 0.19849369 | 4702692.4 |
| 19 | 54.93615 | 5711.5512 | 3726.614 | 0.33354022 | 5016463.1 |
| 20 | 55.46167 | 3726.6144 | 0.000 | 0.39239754 | 5223147.4 |

Actual Cost: 5223147
Expected Cost: 5258727
Improvement: 0.3557954

*Table 2*

This time the actual execution cost outperforms the theoretical optimal by $0.36 a share. Note, accumulated cost often differs from expected cost, as it is only when $\lim_{T \to \infty} \frac{1}{T} \sum_0^T (\varepsilon_t + \eta_t) = 0$.

The Informed model serves to illustrate that in the presence of predictable exogenous variables correlated with stock price, the optimal strategy, though still irrespective of the price-impact from other investor's market orders ($\varepsilon_t$), is now dependent on this series of observed ($X_t$). However, the optimal implementation is more theoretical than practical and without perfect market knowledge, mitigating implementation shortfall becomes a more difficult problem. Without knowing the true market dynamics ($\theta, \gamma, \rho$) and volatility ($\sigma_\varepsilon^2, \sigma_\eta^2$) conditions, the optimal strategy is inefficient, and thus crucial to develop Auto-Regressive estimates of the model parameters ($\hat{\theta}_t, \hat{\gamma}_t, \hat{\rho}_t, \widehat{\sigma_{\eta_t}^2}, \widehat{\sigma_{\varepsilon_t}^2}$).

# The Auto-Regressive Model

Now suppose that UNC has access to some serially correlated market information, however, they are not able to estimate future $(X_t)$ as they do not know how correlated $(\rho)$ the data really is. Additionally, they do not know market information impacts stock price $(\gamma)$ nor do they know the price impact of their own market orders $(\theta)$. How should UNC approach buying ABC's stock so that they acquire ABC shares at the lowest aggregate cost possible, and what cost should they expect to incur given their heedless endeavors? The optimal strategy would be to derive unbiased estimators of the market parameters, such that the guesses are dynamic $(\widehat{\theta}_t, \widehat{\gamma}_t, \widehat{\rho}_t)$, and convergent, $\lim_{T \to \infty} (\widehat{\theta}_t, \widehat{\gamma}_t, \widehat{\rho}_t) = (\theta, \gamma, \rho)$. The later strategy is found to be the Auto-Regressive strategy, a best-execution strategy found assuming an initial lack of knowledge regarding the true market dynamics.

The Auto-Regressive model, more technically known as Additive Permanent Price Impact with Auto-Regressive Estimate of Information, is such that the price-impact of market orders and changes in market information are linear and permanent. In this model there are 7 components, the price-impact from other investors' market orders $(\varepsilon_t)$, the price-impact that our market orders have on execution price $(\theta B_t)$, the price impact of serially correlated market information $(\gamma X_t)$, the true information correlation coefficient $(\rho)$, the dynamic Auto-Regressive estimate of the information correlation coefficient $(\widehat{\rho}_t)$, market information fluctuations $(\eta_t)$, and an Auto-Regressive estimate of variance of its fluctuations $(\widehat{\sigma_\eta^2})$. The $(\varepsilon_t, \eta_t)$ components are given by an arithmetic random walk, and the execution-impact and information impact is an additive linear function of order size $(B_t)$ and information $(X_t)$.

The following equation summarizes the Additive Permanent Price Impact with Auto-Regressive Estimate of Information:

$$P_t = P_{t-1} + \theta B_t + \gamma X_t + \varepsilon_t$$

$$X_t = \rho X_{t-1} + \eta_t$$

$$\varepsilon_t \sim N(0, \sigma_\varepsilon^2), \qquad \eta_t \sim N(0, \sigma_\eta^2)$$

The objective is to find the $B_t$ producing the minimal execution cost strategy, the Bellman equation, and dynamic parameters:

$$V_t = \min_{B_t} \sum_t^T P_t B_t$$

$$a_t = \frac{\widehat{\theta_t}}{2}\left(1 + \frac{1}{T - t + 1}\right), \qquad b_t = \widehat{\gamma_t} + \frac{\widehat{\theta_t}\widehat{\rho_t}}{2}\left(\frac{b_{t-1}}{a_{t-1}}\right),$$

$$c_t = \widehat{\rho_t}^2 c_{t-1} + \frac{\widehat{\rho_t}^2}{4}\left(\frac{b_{t-1}^2}{a_{t-1}}\right), \qquad d_t = d_{t-1} + c_{t-1}\widehat{\sigma_{\eta_t}^2},$$

$$e_t = \frac{1}{T - t + 1}, \qquad f_t = \frac{\widehat{\rho_t}}{2}\left(\frac{b_{t-1}}{a_{t-1}}\right)$$

Solving the Bellman equation by dynamic programming yields the optimal order size and execution cost as:

$$B_t = e_t S_t + f_t X_{t-1}$$

$$V_t = S_t P_t + a_t S_t^2 + b_t S_t X_t + c_t X_t^2 + d_t$$

The impact of an increase in $X_t$ on expected best-execution costs may be measured explicitly by the derivative of the optimal-value function $V_t$ with respect to $X_t : \frac{\partial V_t}{\partial X_t} = b_t B_t + 2c_t X_t$.

Minimizing the Least Squares loss using Fermat's rule and then solving the system of equations yields the estimated the model parameters as:

$$f(\theta, \rho, \gamma) = \sum_1^t (P_k - P_{k-1} - \widehat{\theta_k} B_k - \widehat{\gamma_k} X_k)^2 + \sum_1^t (X_k - \widehat{\rho_t} X_{k-1})^2$$

$$\begin{bmatrix} \widehat{\theta_t} \\ \widehat{\gamma_t} \\ \widehat{\rho_t} \end{bmatrix} = \begin{bmatrix} \sum_1^t B_k^2 & \sum_1^t X_k B_k & 0 \\ \sum_1^t X_k B_k & \sum_1^t B_k^2 & 0 \\ 0 & 0 & \sum_1^t X_k^2 \end{bmatrix}^{-1} \begin{bmatrix} \sum_1^t (P_k - P_{k-1}) B_k \\ \sum_1^t (P_k - P_{k-1}) X_k \\ \sum_1^t X_k X_{k+1} \end{bmatrix}$$

$$\begin{bmatrix} \widehat{\sigma_{\varepsilon_t}^2} \\ \widehat{\sigma_{\eta_t}^2} \end{bmatrix} = \begin{bmatrix} \frac{1}{t-1} \sum_1^t (P_k - P_{k-1} - \widehat{\theta_t} B_k - \widehat{\gamma_t} X_k)^2 \\ \frac{1}{t-1} \sum_1^t (X_k - \widehat{\rho_t} X_{k-1})^2 \end{bmatrix}$$

Solving for $\begin{bmatrix} \widehat{\theta_t} & \widehat{\gamma_t} & \widehat{\rho_t} \end{bmatrix}$ and $\begin{bmatrix} \widehat{\sigma_{\varepsilon_t}^2} & \widehat{\sigma_{\eta_t}^2} \end{bmatrix}$ at each $t$ provides a recursively computed order adjustment factor $\widehat{f_t} = \frac{\widehat{\gamma_t}}{\widehat{\theta_t}(T-t+1)} \sum_1^{T-t} (T-t-k) \widehat{\rho_t}^k$. Further inspection of the optimal order size shows its $\lim_{t \to \infty} E(\widehat{B_t} | X_{t-1}) = \frac{S_t}{T} + \widehat{f_t} X_{t-1}$, meaning the optimal strategy under the Auto-Regressive model is to split shares $(S_0)$ over $(T)$ periods, and adjust by a factor of $\widehat{f_t} X_{t-1}$, which is dependent on the dynamic estimates of $(\widehat{\theta_t}, \widehat{\gamma_t}, \widehat{\rho_t}, \widehat{\sigma_{\varepsilon_t}^2}, \widehat{\sigma_{\eta_t}^2})$ and $t$. Though, as $t \to \infty$, the Auto-Regressive strategy's trading decisions, in practice, will eventually converge to the best-execution of the informed strategy.

Auto-Regressive Numerical Example:

Suppose UNC is subject to the restraint of having to sell 100,000 shares $(S_0)$ in 20 market orders $(T)$, then, given $\theta = 5 \times 10^{-5}$, $\gamma = 5$, $\rho = 0.5$, $\sigma_\varepsilon^2 = (0.125)^2$, and $\sigma_\eta^2 = 0.001$, the Auto-Regressive strategy suggests an optimal split of $\widehat{B_t} \approx 5,000 + \widehat{f_t} X_{t-1}$, and $V_t = \$5,262,500$. Which is initially equivalent to the Naive strategy. However, it should be noted that until the $\lim_{t \to \infty} \widehat{\rho_t} = \rho$, the Auto-Regressive model, will likely yield high execution costs given that

$\frac{1}{t} \sum_k^t |\widehat{X_k} - \widehat{X_{k-1}}| > 0$ for prolonged periods.

Table 3 gives a simulated example of the performance of the Auto-Regressive strategy.

| Period | Price | Shares.Bought | Shares.Remaining | Market.Information | Accumulated.Cost |
|---|---|---|---|---|---|
| 0 | 50.00000 | 0.000 | 100000.000 | 0.00000000 | 0.0 |
| 1 | 50.31700 | 5000.000 | 95000.000 | 0.14530308 | 251585.0 |
| 2 | 50.71744 | 3623.444 | 91376.556 | 0.19631815 | 435356.8 |
| 3 | 50.91375 | 3816.161 | 87560.394 | 0.10994883 | 629651.9 |
| 4 | 51.10695 | 5067.329 | 82493.065 | -0.25956970 | 888627.6 |
| 5 | 51.34499 | 4572.683 | 77920.382 | -0.03178180 | 1123412.0 |
| 6 | 51.47065 | 5059.592 | 72860.790 | -0.02476563 | 1383832.5 |
| 7 | 51.74905 | 5089.751 | 67771.039 | -0.03701625 | 1647222.2 |
| 8 | 51.84179 | 5030.115 | 62740.925 | -0.25105449 | 1907992.4 |
| 9 | 51.92165 | 4180.251 | 58560.674 | -0.20112941 | 2125037.9 |
| 10 | 52.02656 | 3551.352 | 55009.322 | -0.03448234 | 2309802.5 |
| 11 | 52.67161 | 5169.984 | 49839.337 | 0.19758493 | 2582114.0 |
| 12 | 53.15511 | 7044.396 | 42794.942 | 0.08254030 | 2956559.6 |
| 13 | 53.46578 | 6115.230 | 36679.712 | 0.10256641 | 3283637.4 |
| 14 | 53.56272 | 6220.042 | 30459.670 | 0.04277588 | 3616799.7 |
| 15 | 53.78110 | 5482.222 | 24977.447 | -0.19634429 | 3911639.7 |
| 16 | 53.79524 | 3670.963 | 21306.484 | -0.16378855 | 4109120.1 |
| 17 | 53.85467 | 4113.627 | 17192.857 | -0.14423699 | 4330658.1 |
| 18 | 54.13003 | 4777.597 | 12415.260 | -0.08149677 | 4589269.5 |
| 19 | 54.65876 | 5857.945 | 6557.315 | 0.13318090 | 4909457.5 |
| 20 | 55.24812 | 6557.315 | 0.000 | 0.18725913 | 5271736.9 |

```
Actual Cost: 5271737
Expected Cost: 5262500
Improvement: -0.09236862
```

*Table 3*

This time the actual execution cost underperforms the theoretical optimal by -$0.09 a share. Note, accumulated cost often differs from expected cost, as it is only when $T \to \infty$, that

$\lim_{T \to \infty} \frac{1}{T} \sum_0^T (\varepsilon_t + \eta_t) = 0$

The Auto-Regressive model serves to illustrate the periodic differences in the executed strategy and differences in the observed vs expected accumulated execution costs. The implemented strategy using $(\widehat{\theta_t}, \widehat{\gamma_t}, \widehat{\rho_t}, \widehat{\sigma_{\eta_t}^2}, \widehat{\sigma_{\varepsilon_t}^2})$ OLS estimates provide a numerical estimate of the implementation shortfall that investors should expect to incur until model parameter convergence. However, the period $t$ until relative convergence varies depending on the sequence of $(\varepsilon, \eta)$ and thus it is important to analyze the strategy's average performance across many simulations.

## Simulations & Analysis

This section provides a summary overview of the average per period estimated model parameters and accumulated execution costs for 100 simulations of the performance of the Naive, Informed, and Auto-Regressive model under the same sequences of noise $(\varepsilon, \eta)$. The models were evaluated under the following assumptions.

$$S_0 = 100,000, \qquad P_0 = 50, \qquad T = 20$$

$$\theta = 5 \times 10^{-5}, \qquad \gamma = 5, \qquad \rho = 0.5$$

$$\sigma_\varepsilon^2 = (0.125)^2, \qquad \sigma_\eta^2 = 0.001$$

To develop some intuition for these parameters, observe that the no-impact cost of acquiring $S_0$ is $100,000 \times P_0 = \$5\ million$, and the expected full-impact cost is $100,000 \times E[P_0 + \theta S_0 + \gamma X_0 + \varepsilon_0] = \$5.5\ million$, since $E[X_t] = 0$, hence $\theta$ is calibrated to yield an impact of \$500,000 on a 100,000-share block purchase. Additionally, $\sqrt{Var[\gamma X_t]} = \frac{\gamma \sigma_\eta}{\sqrt{1-\rho^2}} = 0.183$, hence the standard deviation of the information component is approximately 18.3 cents (per period). Finally, the standard deviation of the market noise component is calibrated to be 12.5 cents or one stock 'tick' (per period).

Figure 1 shows the convergence of estimate Auto-Regressive model parameters to their true values.

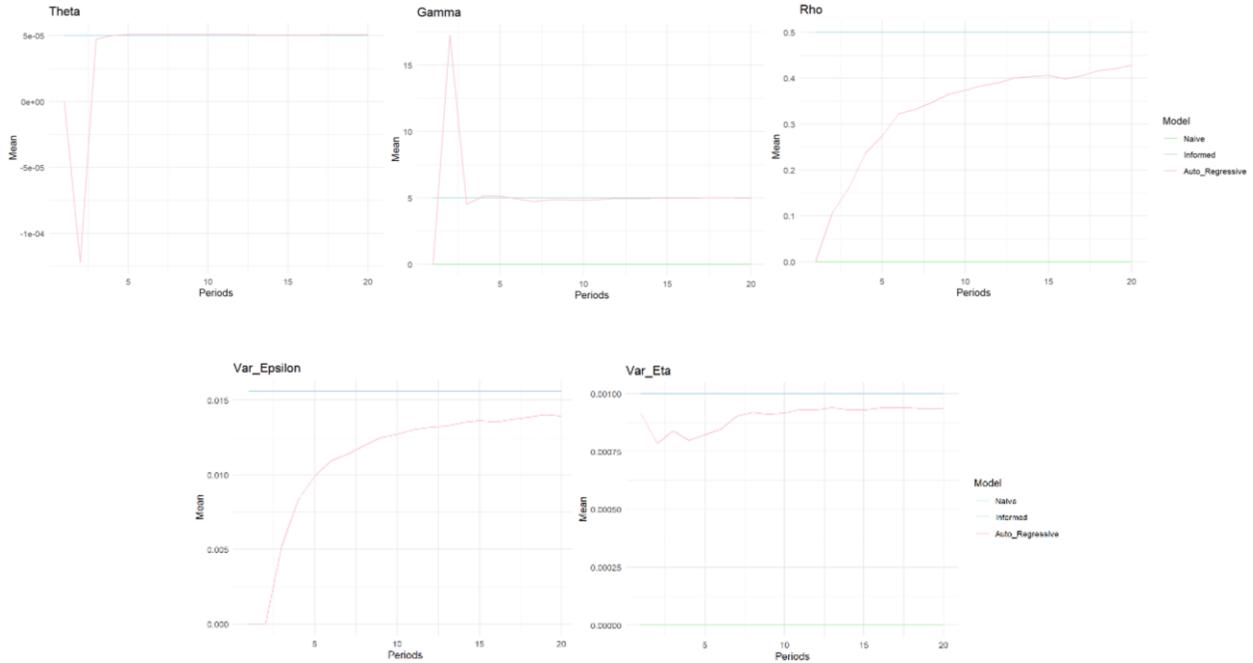

The Auto-Regressive strategy initially deviates from the optimal order size as suggested by Informed strategy, and therefore incurs unnecessary costs without knowledge of the true market parameters. However, $(\hat{\rho}_t, \widehat{\sigma^2_{\eta_t}})$ seem to converge relatively slowly in comparison to $(\hat{\theta}_t, \hat{\gamma}_t, \widehat{\sigma^2_{\varepsilon_t}})$. Note, the convergence to $(\rho, \sigma^2_\eta)$ within a reasonable $t$ is crucial to mitigating implementation shortfall as $\hat{f}_t$ is heavily influences by $(\hat{\rho}_t, \widehat{\sigma^2_{\eta_t}})$. Therefore, it is important to study this impact on aggregate execution costs under the Auto-Regressive in comparison to the Informed strategy and determine the worth of perfect knowledge of initial market dynamics.

Figures 2 & 3 show the average per period market order size with the corresponding per period variance in accumulated execution costs.

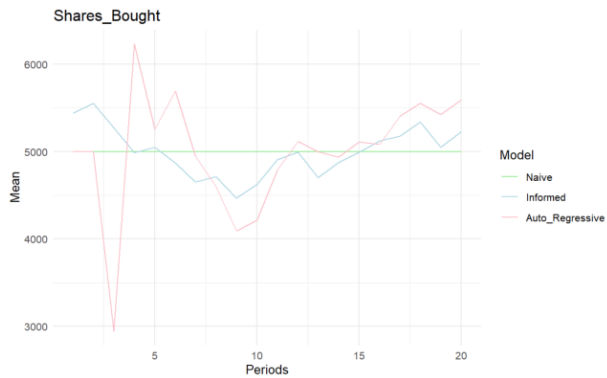

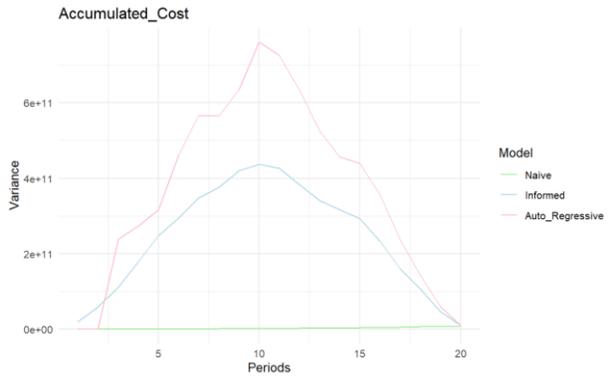

*Figure 2*                                                  *Figure 3*

The figures above show that not only are the Auto-Regressive and Informed strategy's shares bought per period more volatile, but their average accumulated cost appears to be as well. The initial periods appear to have the largest difference in expected order size per model, and the middle periods appear to have the largest difference in the variance of the expectation of the accumulated cost.

Figure 4 shows the average per period accumulated vs expected execution cost across simulations.

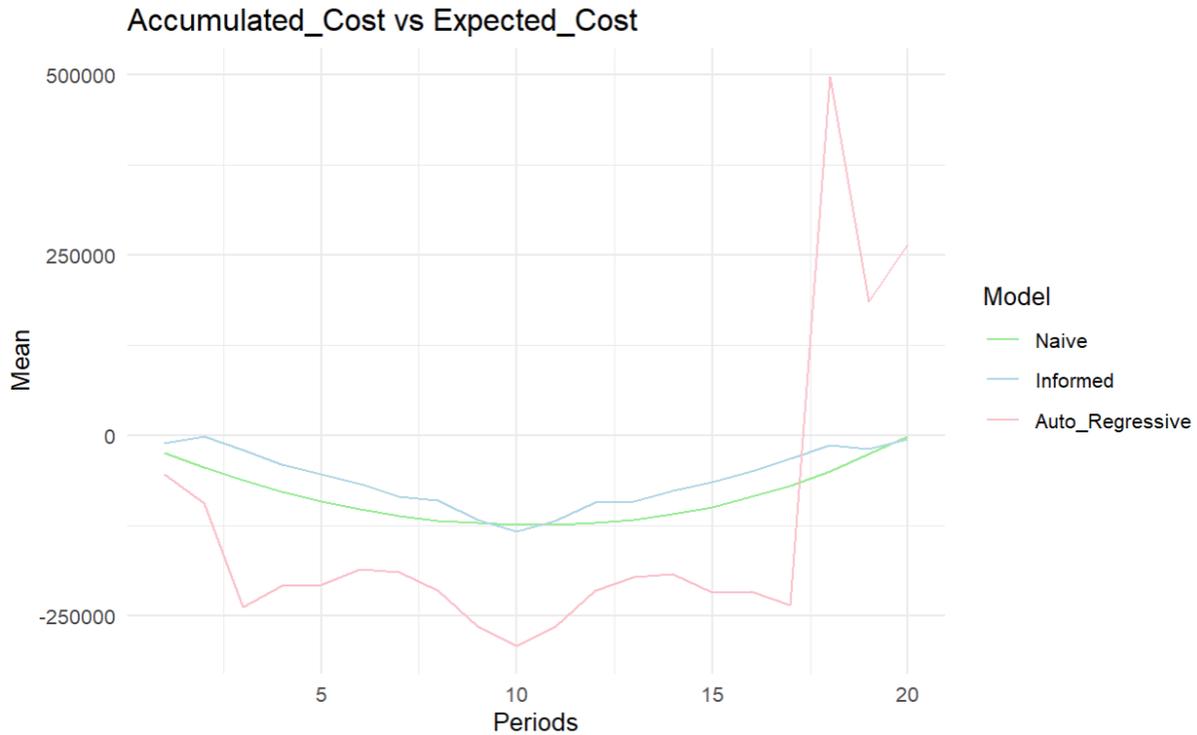

*Figure 4*

In summary, the figures show that in the middle periods, the Informed and Auto-Regressive strategy vary significantly from the Naive strategy, indicating the model's dependence on serially correlated market information. The differences in the final average accumulated execution arise from the strategies' ability to accurately utilize the market information. Though the Informed model has a large variance in the middle periods, its market order size is dynamically calculated according to the market parameters. The Auto-Regressive model, however, must estimate these parameters and its implementation shortfall slowly accumulates until it eventually underperforms compared to both the Naive and Informed strategy. However, the Naive strategy is uniform and its deviations can be summarized by the variance of a $\varepsilon_t$, whereas the Informed and Auto-Regressive strategies are subject to the compound variance of $\varepsilon_t$ and $\eta_t$. It is also important to notice that the change in per period average accumulated cost of the Auto-Regressive strategy eventually converges to that of the Informed strategy. The initial periodic deviations are much more volatile, this is likely due to the initial bias in the estimated model parameters. Until parameter convergence, the implementation shortfall accumulates, thus, it is important to analyze the execution cost impact that $(\hat{\theta}_t, \hat{\gamma}_t, \hat{\rho}_t, \widehat{\sigma_{\varepsilon_t}^2}, \widehat{\sigma_{\eta_t}^2})$ are responsible for.

Figure 5 shows the total accumulated execution cost for each simulation.

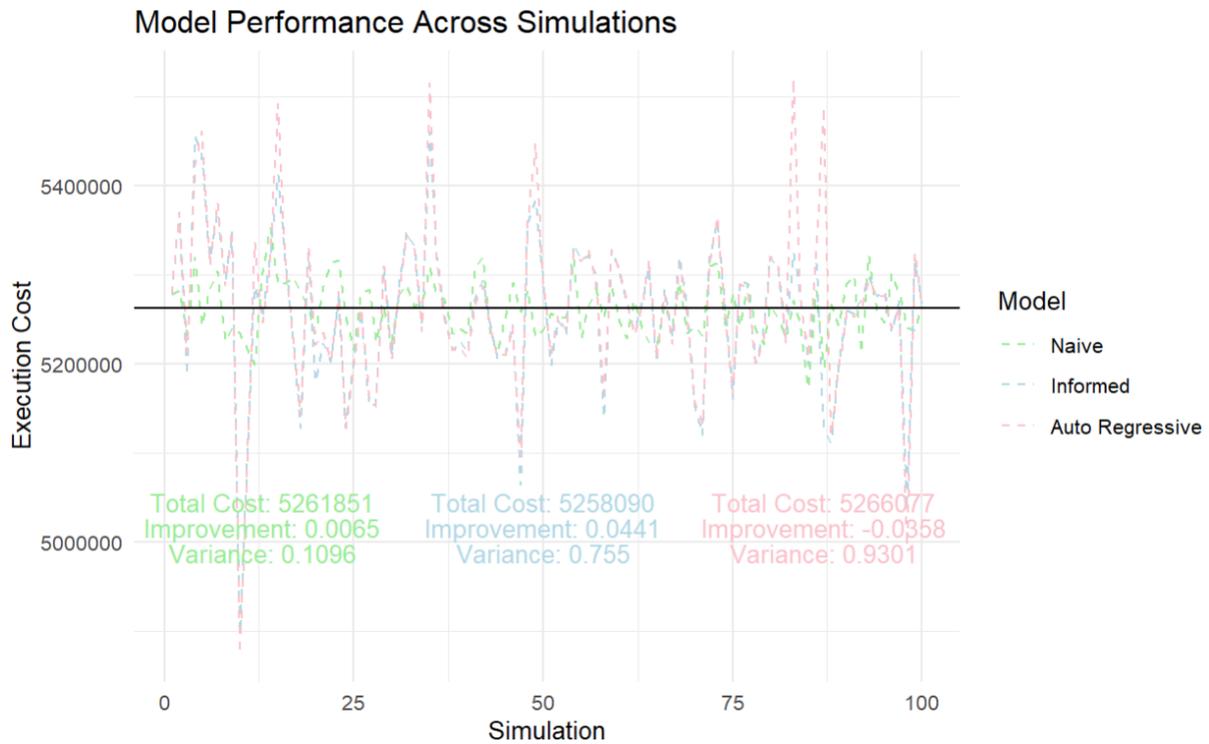

*Figure 5*

The figure above shows each strategies performance across 100 simulated sequences of $(\varepsilon_t, \eta_t)$. Specifically, the Naive, Informed, and Auto-Regressive strategies outperform the Naive expected best-execution cost by around \$0.0065, \$0.0441, and -\$0.0358 a share at each period, respectively.

Additionally, the Naive, Informed, and Auto-Regressive strategies simulated costs show a periodic variance of around $0.1096, $0.7550, and $0.9301 a share, which is significant in comparison to the average improvement, implying that the optimal strategy, though better with reliable market information, may produce very favorable/unfavorable results and should be considered when comparing these models by their risk to reward ratio.

# Conclusion

From my experimentation with the models, the variance of the average improvement across 100 simulations seems to increase exponentially as either the market information relevance factor ($\gamma$) increases, or as the information correlation coefficient ($\rho$) increases and the information variance ($\sigma_\eta^2$) decreases. Logically, these parameter adjustments make sense as ($\gamma$) dictates the maximum expected improvement that the Informed strategy can produce in comparison to the Naive approach. As $\gamma \to 0$ the best strategy becomes Naive, regardless of if serial market information is 100% predictable. Contrastingly, as $\rho \to 0$, the best strategy becomes Naive, regardless of if market price is extremely dependent on market information. Additionally, if $\sigma_\eta^2 \to \infty$, then the best strategy becomes the Naive, as predicting $X_t$ using $\rho$ becomes trivial, and therefore so is $\gamma$.

The Informed strategy dominated Naive approach under many assumptions and restrictions. The Auto-Regressive approach, though showing slightly suboptimal performance, is limited to the same assumptions and restrictions that allow the Informed strategy to dominate. However, the Auto-Regressive approach shows autonomy when prioritizing other price impact parameters ($\hat{\theta}_t$, $\hat{\gamma}_t$, $\widehat{\sigma_{\varepsilon t}^2}$) in the initial periods, as until ($\hat{\rho}_t$, $\widehat{\sigma_{\eta t}^2}$) converge, the Auto-Regressive best execution will emulate the Naive strategy, and if $\hat{\rho}_t$ is very high or $\widehat{\sigma_{\eta t}^2}$ is very low, it will emulate the Informed strategy. Considering that perfect knowledge or market information correlation is unrealistic, and the relative convergence of the Informed vs Auto-Regressive performance as $T \to \infty$. Even though the Auto-Regressive strategy appears to be the inferior model regarding average improvement and variance per share, in retrospect, the increased per share costs may be negligible when considering the price institutional investors typically pay for perfect knowledge of market dynamics. Additionally, without the initial knowledge of underlying market dynamics, the opportunity cost associated with abstinence from placing market orders until convergence of market parameters, may outweigh the average per period costs of implementing the Auto-Regressive strategy. For these reasons, the Auto-Regressive model is in fact a competitive strategy, considering it is the more applicable and realistic in comparison to the models proposed by the literature.

## Improvements

In future experimentation, it would be interesting to increase the degree of autoregression and model $X_t$ as an AR(n) model. Incorporating more periods of the serially correlated market data can only aid the model's ability to predict future price fluctuations and make corresponding adjustments to its buying strategy. Additionally, experimenting with a nonlinear order book where the bid-ask distribution is normally distributed and continuously regenerative around the current market price. This implementation would limit the Informed model's purchasing strategies, making sure that periods of extremely favorable/unfavorable market information are accounted for when considering the exponential price impact of buying large quantities of shares. These improvements should be extended to the Auto-Regressive estimate of all price impact variables, producing multivariate time series model of $\widehat{\theta_t}$, $\widehat{\gamma_t}$, $\widehat{\rho_t}$, $\widehat{\sigma_{\eta_t}^2}$, and $\widehat{\sigma_{\varepsilon_t}^2}$ alongside a more realistic limit order book distribution, to develop a complete understanding of the long-term price-impact in the Auto-Regressive Control of Execution Costs.